\documentclass[letterpaper]{article}
\usepackage{aaai24}

\usepackage{times}  
\usepackage{helvet}  
\usepackage{courier}  
\usepackage{graphicx}
\usepackage{booktabs}
\usepackage{comment}
\usepackage{enumerate}
\usepackage{caption}
\usepackage{natbib}
\usepackage[hyphens]{url}  
\usepackage{graphicx} 
\usepackage{hyphenat}
\usepackage{subfig}
\usepackage{enumitem}

\usepackage{amsmath, amsthm, amssymb, mathtools}
\usepackage{array, multirow, multicol}
\usepackage{xcolor}

\newcommand{\ignore}[1]{}

\usepackage{xspace}
\usepackage{amsmath}
\usepackage{amsfonts}
\usepackage{url}
\usepackage{marvosym}
\usepackage{ifthen}
\theoremstyle{plain}

\newcommand{\chatoDisplayMode}[1]{#1}

\definecolor{MyRed}{rgb}{0.6,0.0,0.0} 
\definecolor{MyBlack}{rgb}{0.1,0.1,0.1} 
\newcommand{\inred}[1]{{\color{MyRed}\sf\textbf{\textsc{#1}}}}
\newcommand{\frameit}[2]{
  \begin{center}
  {\color{MyRed}
  \framebox[.9\columnwidth][l]{
    \begin{minipage}{.85\columnwidth}
    \inred{#1}: {\sf\color{MyBlack}#2}
    \end{minipage}
  }\\
  }
  \end{center}
}

\newcommand{\note}[2][]{\chatoDisplayMode{\def\@tmpsig{#1}\frameit{{\Pointinghand} Note}{#2\ifx \@tmpsig \@empty \else \mbox{ --\em #1}\fi}}}
\newcommand{\todo}[2][]{\chatoDisplayMode{\def\@tmpsig{#1}\frameit{{\Writinghand} To-do}{#2\ifx \@tmpsig \@empty \else \mbox{ --\em #1}\fi}}}

\newcommand{\abbrevStyle}[1]{#1}

\newcommand{\ie}{\abbrevStyle{i.e.}\xspace}
\newcommand{\eg}{\abbrevStyle{e.g.}\xspace}
\newcommand{\cf}{\abbrevStyle{cf.}\xspace}

\newcommand{\vs}{\abbrevStyle{vs.}\xspace}
\newcommand{\etc}{\abbrevStyle{etc.}\xspace}

\newcommand{\Secref}[1]{Sec.~\ref{#1}}

\newcommand{\xhdr}[1]{\vspace{1.7mm}\noindent{{\bf #1.}}}

\newcommand{\textcite}[1]{\citeauthor{#1} \shortcite{#1}}

\newcommand{\hide}[1]{}

\makeatletter
\newcommand{\iffont}[2]{\ifthenelse{\equal{\f@family}{#1}}{#2}{}}
\makeatother

\iffont{ptm}{
  \usepackage{mathptmx}

  \DeclareSymbolFont{greek}{OML}{cmm}{m}{n}
  \DeclareMathSymbol{\alpha}{\mathalpha}{greek}{"0B}
  \DeclareMathSymbol{\beta}{\mathalpha}{greek}{"0C}
  \DeclareMathSymbol{\gamma}{\mathalpha}{greek}{"0D}
  \DeclareMathSymbol{\delta}{\mathalpha}{greek}{"0E}
  \DeclareMathSymbol{\epsilon}{\mathalpha}{greek}{"0F}
  \DeclareMathSymbol{\zeta}{\mathalpha}{greek}{"10}
  \DeclareMathSymbol{\eta}{\mathalpha}{greek}{"11}
  \DeclareMathSymbol{\theta}{\mathalpha}{greek}{"12}
  \DeclareMathSymbol{\iota}{\mathalpha}{greek}{"13}
  \DeclareMathSymbol{\kappa}{\mathalpha}{greek}{"14}
  \DeclareMathSymbol{\lambda}{\mathalpha}{greek}{"15}
  \DeclareMathSymbol{\mu}{\mathalpha}{greek}{"16}
  \DeclareMathSymbol{\nu}{\mathalpha}{greek}{"17}
  \DeclareMathSymbol{\xi}{\mathalpha}{greek}{"18}
  \DeclareMathSymbol{\pi}{\mathalpha}{greek}{"19}
  \DeclareMathSymbol{\rho}{\mathalpha}{greek}{"1A}
  \DeclareMathSymbol{\sigma}{\mathalpha}{greek}{"1B}
  \DeclareMathSymbol{\tau}{\mathalpha}{greek}{"1C}
  \DeclareMathSymbol{\upsilon}{\mathalpha}{greek}{"1D}
  \DeclareMathSymbol{\phi}{\mathalpha}{greek}{"1E}
  \DeclareMathSymbol{\chi}{\mathalpha}{greek}{"1F}
  \DeclareMathSymbol{\psi}{\mathalpha}{greek}{"20}
  \DeclareMathSymbol{\omega}{\mathalpha}{greek}{"21}
  \DeclareMathSymbol{\varepsilon}{\mathalpha}{greek}{"22}
  \DeclareMathSymbol{\vartheta}{\mathalpha}{greek}{"23}
  \DeclareMathSymbol{\varpi}{\mathalpha}{greek}{"24}
  \DeclareMathSymbol{\varrho}{\mathalpha}{greek}{"25}
  \DeclareMathSymbol{\varsigma}{\mathalpha}{greek}{"26}
  \DeclareMathSymbol{\varphi}{\mathalpha}{greek}{"27}
  \DeclareSymbolFont{otone}{OT1}{cmr}{m}{n}
  \DeclareMathSymbol{\Gamma}{\mathalpha}{otone}{0}
  \DeclareMathSymbol{\Delta}{\mathalpha}{otone}{1}
  \DeclareMathSymbol{\Theta}{\mathalpha}{otone}{2}
  \DeclareMathSymbol{\Lambda}{\mathalpha}{otone}{3}
  \DeclareMathSymbol{\Xi}{\mathalpha}{otone}{4}
  \DeclareMathSymbol{\Pi}{\mathalpha}{otone}{5}
  \DeclareMathSymbol{\Sigma}{\mathalpha}{otone}{6}
  \DeclareMathSymbol{\Upsilon}{\mathalpha}{otone}{7}
  \DeclareMathSymbol{\Phi}{\mathalpha}{otone}{8}
  \DeclareMathSymbol{\Psi}{\mathalpha}{otone}{9}
  \DeclareMathSymbol{\Omega}{\mathalpha}{otone}{10}
  \DeclareSymbolFont{syms}{OML}{cmm}{m}{it}
  \DeclareMathSymbol{\partial}{\mathord}{syms}{"40}
  \DeclareMathAlphabet{\mathbold}{OML}{cmm}{b}{it}
  \DeclareSymbolFont{largesymbols}{OMX}{cmex}{m}{n}

}

\title{Orphan Articles: The Dark Matter of Wikipedia}
\author {
    Akhil Arora\textsuperscript{\rm 1},
    Robert West\textsuperscript{\rm 1}\thanks{Robert West is a Wikimedia Foundation Research Fellow.},
    Martin Gerlach\textsuperscript{\rm 2}
}
\affiliations {
    \textsuperscript{\rm 1}EPFL\\
    \textsuperscript{\rm 2}Wikimedia Foundation\\
    akhil.arora@epfl.ch, robert.west@epfl.ch, mgerlach@wikimedia.org
}

\begin{document}

\maketitle

\begin{abstract}
With 60M articles in more than 300 language versions, Wikipedia is the largest platform for open and freely accessible knowledge. 
While the available content has been growing continuously at a rate of around 200K new articles each month, very little attention has been paid to the discoverability of the content. 
One crucial aspect of discoverability is the integration of hyperlinks into the network so the articles are visible to readers navigating Wikipedia. 
To understand this phenomenon, we conduct the first systematic study of \emph{orphan articles}, which are articles without any incoming links from other Wikipedia articles, across 319 different language versions of Wikipedia. 
We find that a surprisingly large extent of content, roughly 15\% (8.8M) of all articles, is de facto invisible to readers navigating Wikipedia, and thus, rightfully term orphan articles as the \emph{dark matter} of Wikipedia. 
We also provide causal evidence through a \emph{quasi-experiment} that adding new incoming links to orphans (de-orphanization) leads to a statistically significant increase in their visibility in terms of the number of pageviews. 
We further highlight the challenges faced by editors for de-orphanizing articles, demonstrate the need to support them in addressing this issue, and provide potential solutions for developing automated tools based on cross-lingual approaches. 
Overall, our work not only unravels a key limitation in the link structure of Wikipedia and quantitatively assesses its impact but also provides a new perspective on the challenges of maintenance associated with content creation at scale in Wikipedia.
\end{abstract}

\section{Introduction}
Wikipedia is the largest multi-lingual platform on the Internet for open and freely accessible knowledge. 
As of November 2022, Wikipedia comprised 60M articles across 319 different language versions, and it has since been growing at a rapid rate of around 200K articles per month. 
In fact, in order to bridge knowledge gaps~\cite{redi2020taxonomy}, there have been a plethora of efforts to systematically add content that is currently absent, \eg, formation of organized groups such as Wiki Women in Red~\cite{wiki_women_in_red} to add articles about women~\cite{vitulli2018writing}, development of automatic tools such as Project Quicksilver~\cite{quicksilver} to surface missing articles by generating a list of people who are missing from Wikipedia based on news, or translation of existing articles into other languages~\cite{wulczyn2016growing}. 
These initiatives have been extremely successful---for example, Wikipedia's content translation tool~\cite{content_translation} has helped to create more than 1M new articles~\cite{ozurumba2021content}. 
As a result, one of the main challenges is how to maintain this ever-increasing volume of content. 
Specifically, it is crucial to properly integrate new articles into the existing network structure. 
In fact, hyperlinks play a crucial role in the encyclopedia and editors have developed a dedicated guideline to ``build the web'' in English Wikipedia’s manual of style~\cite{wiki_manual_of_style}, primarily to enable readers to access relevant information on other Wikipedia pages easily. 
While the largest share of traffic to Wikipedia comes from search engines, a substantial fraction ($38\%$) of pageviews result from traffic via internal hyperlinks~\cite{piccardi2023large}.

Thus, it is problematic if existing articles are not integrated into the network structure because they will suffer from a lack of visibility to readers. 
In addition, the lack of visibility reflects structural biases such as the gender gap~\cite{beytia2022visibility}. 
For example, the visibility of biographies about women is systematically lower than for biographies about men ~\cite{wagner2015man,wagner2016women}. 
Different community-driven campaigns, which have been successfully adding and improving content about women, have been shown to be less successful at addressing structural biases that limit their visibility~\cite{langrock2022gender}. 
Previous research demonstrated in a quasi-experiment a spill-over effect of attention in Wikipedia~\cite{kummer2014spillovers,zhu2020content} suggesting a causal relation that visibility can be improved by adding relevant incoming links to articles. 
This provides evidence that the lack of visibility can be improved by suitable interventions.

In this work, we explore the question of the lack of visibility of articles in more than 300 language versions of Wikipedia. 
We specifically focus on so-called \emph{orphan articles}, which are defined as articles that do not have any incoming links from other articles in the main namespace of Wikipedia.\footnote{\url{https://en.wikipedia.org/wiki/Wikipedia:Orphan}} 
These articles are of particular interest since they are de facto invisible for readers navigating hyperlinks in Wikipedia. 
Specifically, we aim to address the following research questions:
\begin{itemize}
    \item \textbf{RQ1}: What are the key characteristics of orphan articles? (\Secref{sec:characterizing-orphans})
    \item \textbf{RQ2}: Does adding incoming links (de-orphanization) increase the visibility of orphan articles? (\Secref{sec:visibility-orphans})
    \item \textbf{RQ3}: What is the current state of de-orphanization and what are the potential ways to improve it? (\Secref{sec:challenges-deorphanization})
\end{itemize}

To answer the aforementioned research questions, we conduct the first systematic study on orphan articles in Wikipedia and show that orphans make up a surprisingly large fraction of articles. We also establish causal evidence through quasi-experiments that orphan articles are significantly less visible than non-orphan articles. We then describe the challenges faced by editors in addressing this issue and sketch potential solutions to develop models to support their efforts, demonstrating the opportunities for using our insights in future works. Together, these results provide a new perspective on maintenance costs associated with content creation and challenges in making existing knowledge discoverable.

\section{Related Work}
In this section, we review existing works that overlap closely with our study. For additional related work, please see Appendix~\ref{app:rwork}.

\xhdr{Orphans articles in Wikipedia}
The English Wikipedia contains an information page about orphan articles~\cite{wiki_orphans}, 
which states that ``these pages can still be found by searching Wikipedia, but it is preferable that they can also be reachable by links from related pages; it is therefore helpful to add links from other suitable pages with similar or related information.'' 
Moreover, according to the manual of style~\cite{wiki_manual_of_style}, de-orphanizing articles is an important aspect of ``building the web'', as hyperlinks are crucial for helping readers in conveniently finding related information while reading Wikipedia. 
Even the guide for creating new pages suggests to ``Provide internal links to the article from other pages''.\footnote{\url{https://en.wikipedia.org/wiki/Help:Drawing_attention_to_new_pages}}
Editors use a maintenance template~\cite{template_orphan} to mark orphan articles with a note visible at the top of the article and organize them in specific categories. 
A group of editors from WikiProject Orphanage~\cite{wikiproject_orphanage} are ``dedicated to clearing up the immense backlog of orphaned articles'' and provide suggestions for how to de-orphanize articles. 
For this, they have a set of tools at their hand such as \emph{findlink} (mentioned in the hat-note of each orphan), which suggests new links from where to link an article based on string matches of the page-title~\cite{findlink}. 
However, despite the organized efforts, in English Wikipedia, the number of articles tagged with an orphan template has been decreasing very slowly from a peak of 140K in 2017 to around 80K in 2023~\cite{wiki_orphan_category}. 
In opposition to orphan articles (no incoming links), there are the so-called dead-end articles, which are articles that contain no outgoing links to other Wikipedia articles. 
Similarly, these articles are marked with a maintenance template~\cite{template_deadend} but for English Wikipedia, the number of affected articles is in the low single digits.

\xhdr{Spillover effect}
Different recent studies have demonstrated a so-called spillover effect in Wikipedia~\cite{kummer2014spillovers,kummer2018attention,zhu2020content}.
These are based on quasi-experiments suggesting a causal effect of newly added incoming links to the attention received by articles. 
For example, \cite{zhu2020content} compared a ``treatment'' group of articles edited through organized campaigns with a ``control'' group of articles that did not receive any edits, finding a significant increase in the number of pageviews for articles that were newly linked from the treated articles but that themselves were neither in the control or treatment group. 
However, none of the aforementioned studies investigated orphan articles specifically.

\xhdr{Knowledge gaps and visibility}
Wikipedia and its sister projects such as Wikimedia Commons, Wikidata, or Wiktionary, suffer from a wide range of knowledge gaps~\cite{redi2020taxonomy}. 
For example, the content gender gap refers to the fact that only 15–20\% of biography articles in Wikipedia are about women~\cite{konieczny2018gender}. 
This gap has been confirmed in countless studies also taking into account more nuanced metrics such as notability~\cite{tripodi2021ms} and has also been confirmed beyond content for the population of editors~\cite{hill2013wikipedia,ford2017anyone} and readers~\cite{johnson2021global}.
One often overlooked aspect, however, has been raised about biases in the visibility of already existing content~\cite{beytia2022visibility}. 
Anecdotally, it has been reported that deletion of biography articles is more common if the subject is not yet mentioned on other Wikipedia articles~\cite{vitulli2018writing}. 
Several studies documented how articles about women are systematically less visible than articles about men using proxies such as PageRank centrality~\cite{wagner2015man,wagner2016women}. 
This is especially interesting in view of studies showing that organized campaigns are successful at adding content about women that would otherwise be missing, but are less successful at addressing structural biases that limit the visibility of women-focused content as addressing these biases directly is non-trivial~\cite{langrock2022gender}. Also, there is evidence that using existing tools for recommending links to support editors in addressing this problem could actually reinforce those biases~\cite{ferrara2022link}.

Complementary to all the aforementioned works, with the primary focus on visibility, orphan articles provide a more nuanced approach to measuring knowledge gaps in Wikimedia projects~\cite{redi2020taxonomy}.

\xhdr{Cross-lingual approaches in Wikipedia}
With more than 300 active language versions, Wikipedia is an intrinsically multilingual project. 
While there are community-created lists of vital articles,\footnote{\url{https://meta.wikimedia.org/wiki/List_of_articles_every_Wikipedia_should_have}}, \ie, articles that every Wikipedia should have, it has been found that there are substantial differences between different language versions~\cite{hecht2010tower,bao2012omnipedia}. 
Thanks to the efforts of, among others, multilingual editors~\cite{hale2014multilinguals}, content has been shown to propagate from one language version to another~\cite{valentim2021tracking,yoon2022quantifying}.
This has also been the motivation for leveraging content translation systematically in order to grow the different language versions of Wikipedia on the level of articles~\cite{wulczyn2016growing}, sections~\cite{piccardi2018structuring}, or section titles~\cite{section_alignment}. 
Specifically, the model for recommending articles for translation has been developed into a tool by the Wikimedia Foundation~\cite{laxstrom2015content}.
This tool supports editors to translate articles from one language into another by providing a first draft of the article using automatic translation~\cite{content_translation}. 
This approach has been extremely successful, with over 1M translated articles as of 2021~\cite{ozurumba2021content}. 
Along similar lines, one recent study proposed to improve the overall inter-connectivity among articles by taking advantage of existing links in other language versions~\cite{lotkowski2017automatic}; however, the work considers only one specific case translating any possible link from English to Scots Wikipedia.

\begin{figure}[t]
    \centering
    \includegraphics[width=0.7\linewidth]{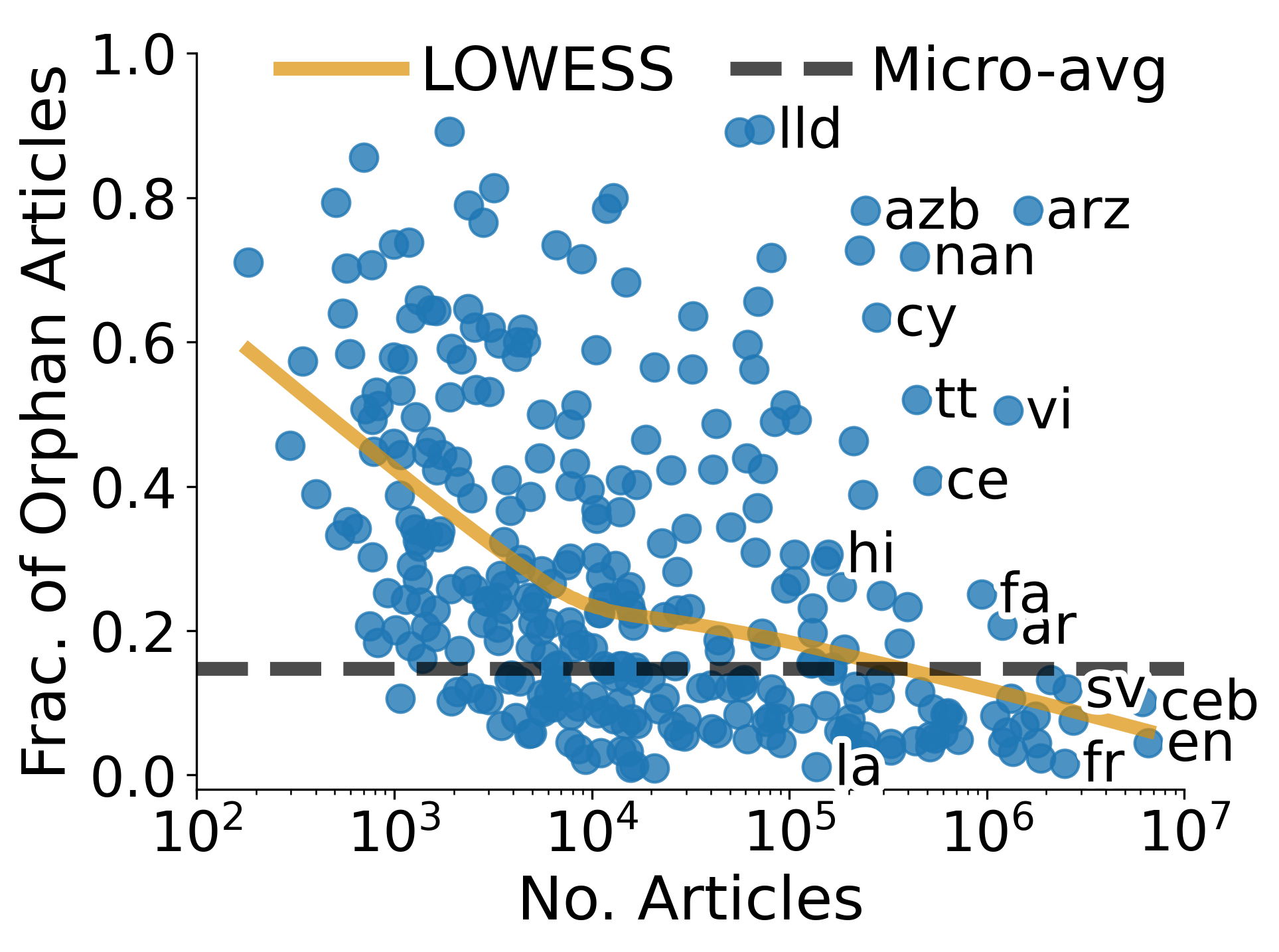}
    \caption{Analyzing the extent of orphan articles across all Wikipedia language versions.}
    \label{fig:orphans_per_wiki}
\end{figure}

\section{Data and Resources}
\label{sec:data}
In this section, we describe in detail the datasets used for studying orphan articles in Wikipedia. 
We consider 319 different language versions of Wikipedia and collect data spanning 7 monthly snapshots ranging from August 2022 to February 2023. 
Unless stated otherwise, the results presented in this paper are based on the monthly snapshot of November 2022. For other snapshots, the results portrayed similar trends, and are therefore omitted. 
All the publicly accessible resources (data, descriptive statistics, and code) required to reproduce the analyses in this paper are available at~\url{https://github.com/epfl-dlab/wikipedia-orphans}.

\xhdr{Wikipedia hyperlink network} For each language version, we construct its `directed' hyperlink network by leveraging the \texttt{pagelinks} table, which tracks all internal links among Wikipedia articles and is available as a SQL dump~\cite{wikipedia_xmldumps} released by Wikipedia on a monthly basis. Note that we resolve redirects~\cite{hill2014consider} and only consider links between articles in the main namespace\footnote{Articles in Wikipedia are grouped into collections called `namespaces', which differentiate between their purpose at a high level. For details please see~\url{https://w.wiki/6hoy}.} of Wikipedia. Specifically, we use the dumps released on the first of each month, \eg, for November 2022, we use the dump dated `2022-11-01' to extract a total of 60M articles and 3.5B links across 319 language versions. Additionally, using the Wikidata dump~\cite{wikidata_dumps} dated `2023-02-26', each article was appropriately mapped to its corresponding unique language-agnostic Wikidata identifier (QID), which further facilitates matching articles across languages.

\xhdr{Orphan-articles data} This data consists of orphan articles in Wikipedia, which are articles with no incoming links from any other main namespace articles in the same language version of Wikipedia. This data is used primarily in Sec.~\ref{sec:characterizing-orphans}.

\xhdr{De-orphanizing-links data} This data consists of new incoming links added to orphan articles. Specifically, for a given month, \eg, November 2022, we obtain the added links by computing the set difference between links existent in Wikipedia in December and November 2022, respectively. Next, to obtain the de-orphanizing links we restrict ourselves to added links with orphan articles from November 2022 as the target. This data is used primarily in Sec.~\ref{sec:visibility-orphans}.

\begin{figure}[t]
    \centering
    \includegraphics[width=0.7\linewidth]{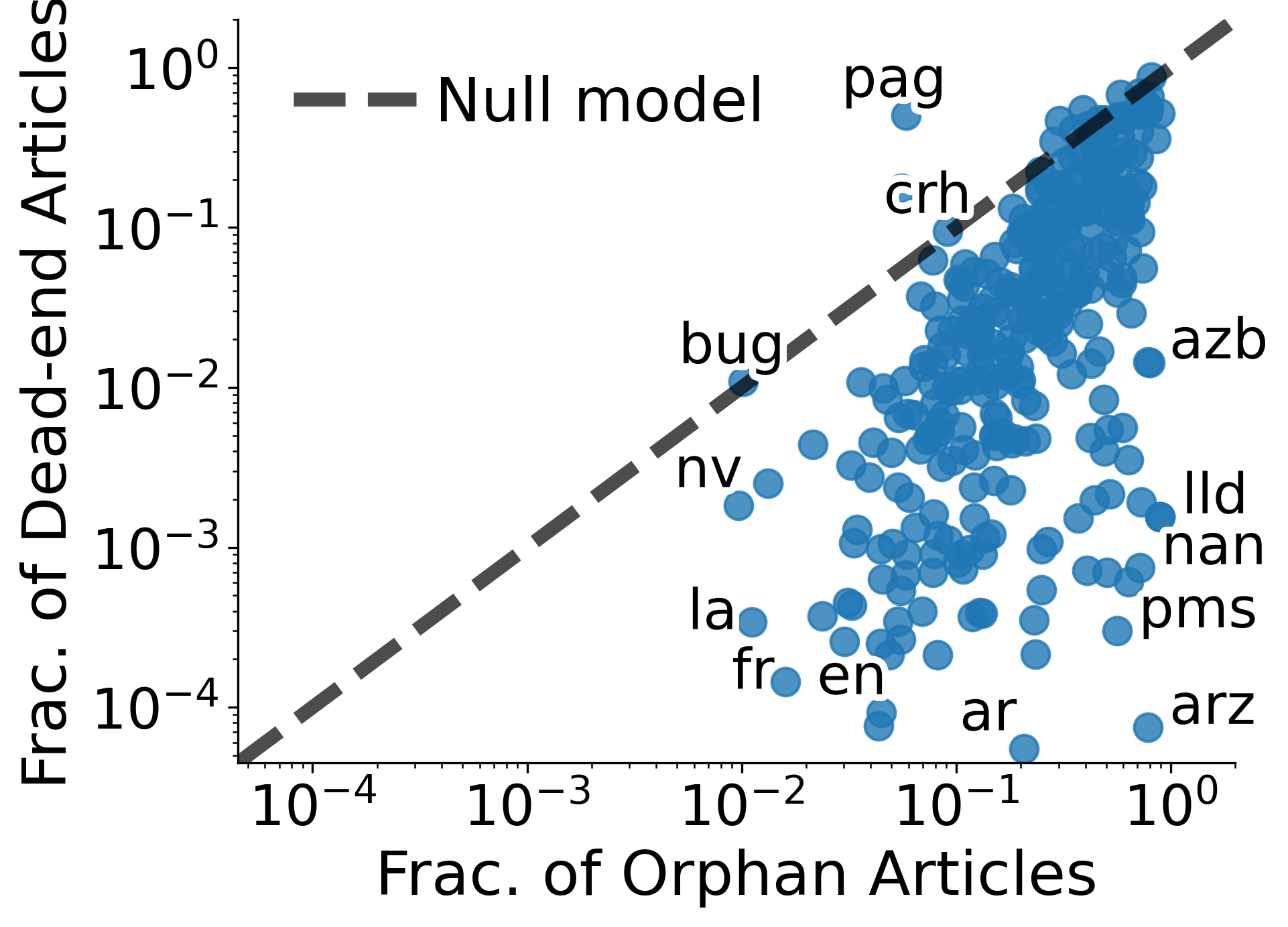}
    \caption{Comparing the extent of orphans with that of of dead-end articles across all Wikipedia language versions.}
    \label{fig:orphans_vs_deadend}
\end{figure}

\begin{figure*}[t]
\centering
\includegraphics[width=0.99\linewidth]{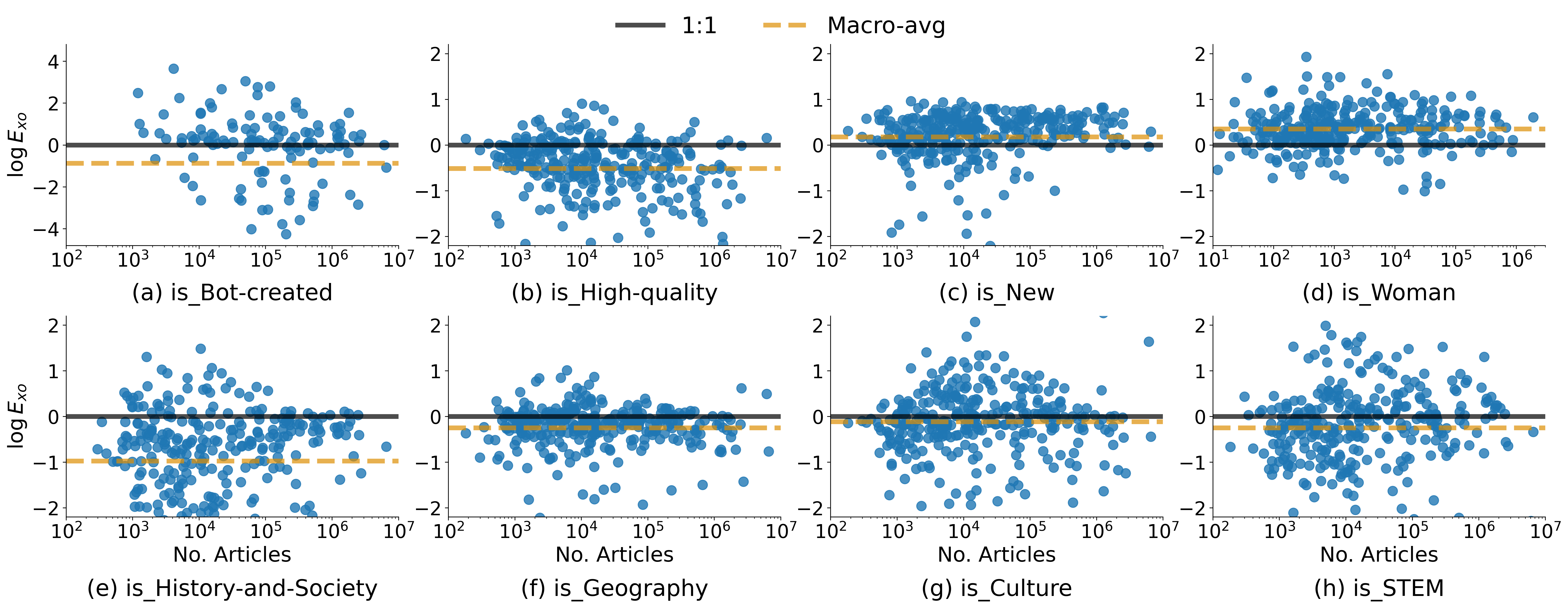}
\caption{Characterizing orphans based on article features in all Wikipedia language versions. For a given feature and language, points above the $\mathbf{y=0}$ line indicate an over-representation of the feature among orphans in that language.}
\label{fig:characterizing_orphans}
\end{figure*}

\xhdr{Wikipedia article features} For each article, we extract the following features: topic, quality, time since creation (age), whether it was created by a bot, the gender (for biography articles), and pageviews.
\begin{itemize}[leftmargin=*, noitemsep]
\item \textbf{Topics}: We use the language-agnostic topic model developed by~\cite{LanguageAgnosticORES}, which assigns topic labels to articles based on the taxonomy~\cite{wikitax} developed with inputs from the Wikipedia editor community~\cite{wp_council}. We use the 4 top-level topic labels from the taxonomy, namely `Culture', `Geography', `History and Society', and `STEM'. For each topic label, the model predicts the probability for an article to be assigned to that topic, and the label is assigned only if the predicted probability is $>0.5$.
\item \textbf{Quality}: Article quality is computed using the language-agnostic quality model by~\cite{language_agnostic_quality}, which uses features such as article-length, number of links, sections, references, \etc, to obtain a score between $[0,1]$. 
\item \textbf{Age}: For each article, we extract its creation timestamp from its revision history, and represent its age using the UNIX timestamp format.
\end{itemize}

\section{Characterizing Orphan Articles}
\label{sec:characterizing-orphans}
In this section, we assess the extent of orphan articles, contrast it with the extent of dead-end articles, and characterize orphan articles based on different article features in all language versions of Wikipedia.

\xhdr{Extent of orphans} Counting the number of orphan articles, we find that the fraction of orphan articles is surprisingly large (Fig.~\ref{fig:orphans_per_wiki}): out of the total 60M articles across all the 319 Wikipedia language versions, 8.8M (14.7\%) are orphan articles. 
This observation is not driven by only a few outliers but is consistent across (almost) all language versions of Wikipedia: there are more than 100 Wikipedia language versions with at least 30\% orphan articles. 
As portrayed by the LOWESS regression fit~\cite{lowess}, smaller Wikipedia language versions tend to have a higher fraction of orphans; yet larger Wikipedia language versions can also have above-average orphan-rates. 
For example, among the 20 largest Wikipedia language versions, we find that Egyptian Arabic (arz, 78\%), Vietnamese (vi, 50\%), Persian (fa, 25\%), and Arabic (ar, 21\%) portray high orphan-rates. 
In relative terms, English Wikipedia is an outlier with only 5\% orphans, however, this still corresponds to more than 300K articles.

\xhdr{Comparison with dead-end articles} In comparison to orphan articles (no incoming links), at 300K ($\sim$ 0.5\% of all articles), dead-end articles (no outgoing links) can be considered virtually non-existent (Fig.~\ref{fig:orphans_vs_deadend}). 
For almost all Wikipedia language versions, we find that the fraction of dead-ends is often (at least) an order of magnitude lower than the fraction of orphans. 
For example, Egyptian Arabic (arz) possesses 1.25M orphans (78\%) but only 121 dead-ends (0.007\%). 
We thus find that the problem of orphan articles is very \emph{distinct from and of much larger scope} than the problem of dead-end articles. This is perhaps intuitive: while the issue of dead-end articles can be addressed by editing the respective article itself, orphan articles can only be addressed by (identifying and) editing other articles.

\xhdr{Characterizing orphans} To better understand which types of articles are found more commonly among orphans (such as whether the article is about a specific topic), we perform a characterization of orphan articles based on the article features described in Sec.~\ref{sec:data}.
For a given feature ($x$), we calculate how many of the orphan articles ($o$) have that feature, \ie, the conditional probability $P(x|o)$. 
By comparing $P(x|o)$ with the overall propensity of the feature $x$ among all articles ($P(x)$), the ratio $\log E_{xo} = P(x|o)/P(x)$ shows whether feature $x$ is over-represented ($\log E_{xo}>0$) or under-represented ($\log E_{xo}<0$) among orphan articles. 

Note that for the purpose of this analysis we require binary article features. While most article features are binary by construction, we binarize the numeric features `age' and `quality' by partitioning the set of articles in each language into two groups---old \vs new and high \vs low quality, respectively---using their median value. We investigated the following article features and whether they are over- or under-represented among orphans (Fig.~\ref{fig:characterizing_orphans}). 

\begin{figure*}[t]
    \centering
    \subfloat[Correlation: Non-orphans \vs Orphans]{
        \includegraphics[width=0.34\linewidth]{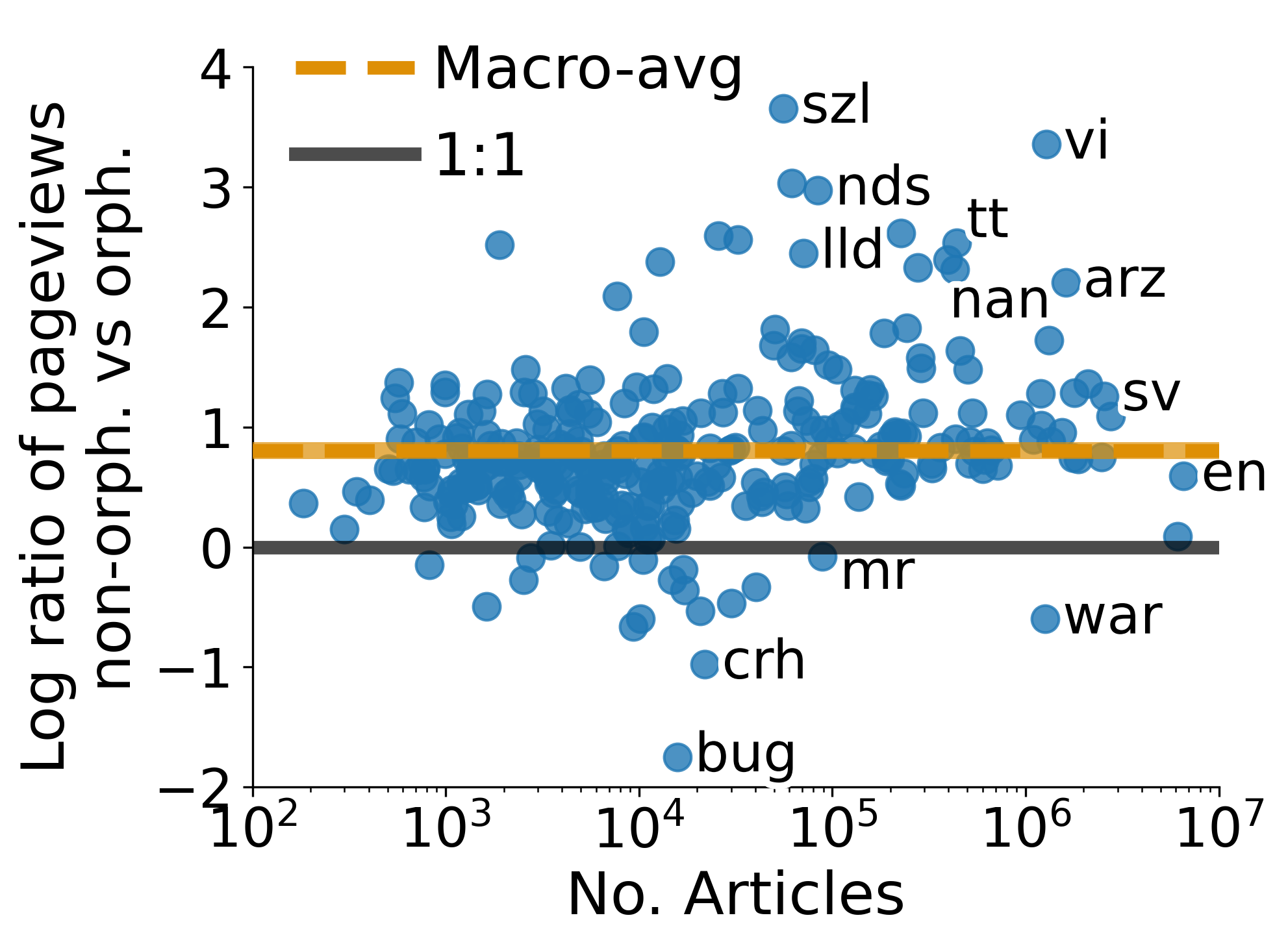}
        \label{fig:visibility_orphans_vs_non-orphans}
    }
    \subfloat[Quasi-experiment: Treated (orph. $\rightarrow$ de-orph.) \vs Control (orph. $\rightarrow$ orph.)]{
        \includegraphics[width=0.57\linewidth]{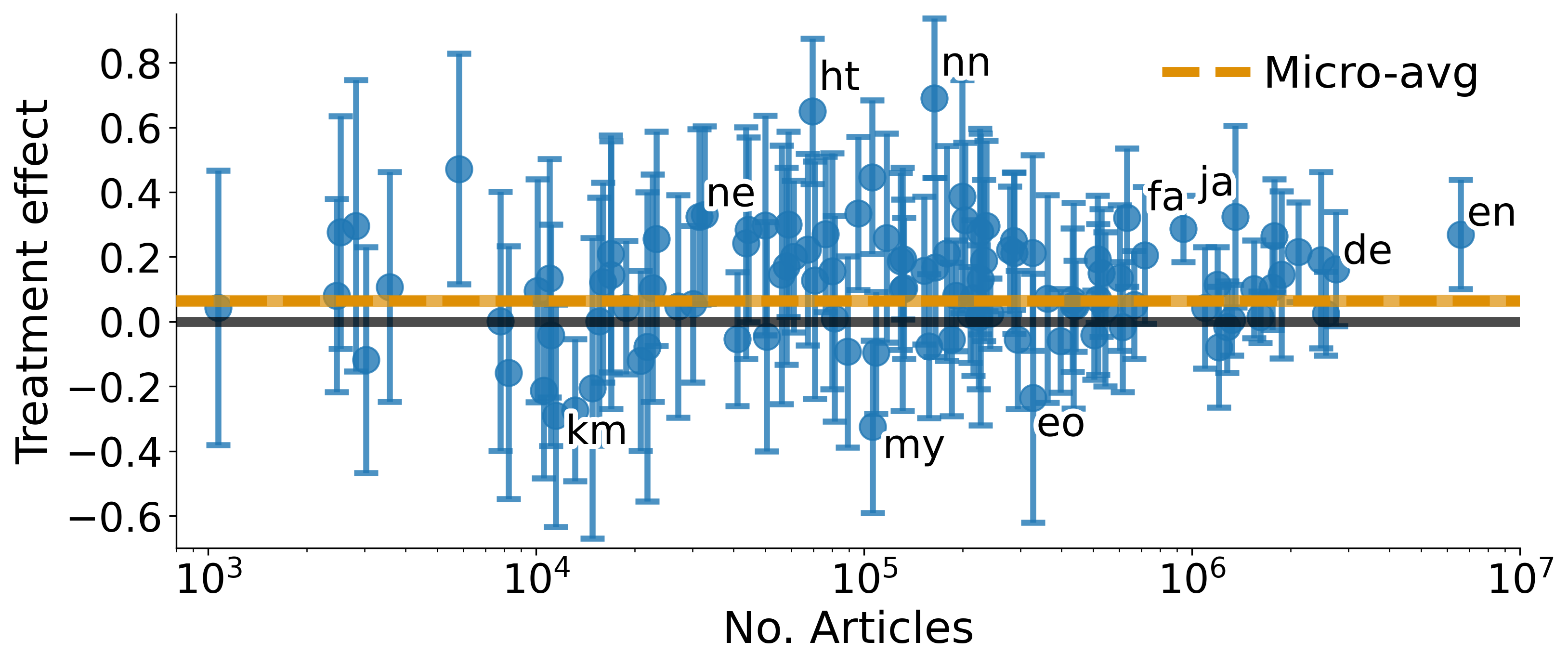}
        \label{fig:lang_specific_treatment_effect}
    }
    \caption{Comparing the pageviews received by orphan and non-orphan articles across all Wikipedia language versions. The error bars denote 95\% CIs, and have been omitted from (a) as they were small and therefore impacting readability.}
    \label{fig:visibility}
\end{figure*}

\textbf{Bot-created} articles are under-represented among orphans. However, the variation is large and there is a considerable number of Wikipedia language versions for which bot-created articles are substantially over-represented. 
For example, among the 20 largest Wikipedia language versions, we find that in Italian (it) Wikipedia (similar trends observed for Chinese (zh) and Portugese (pt)), $P(\mathrm{bot}|o) = 0.19$, which is much larger than the overall fraction of bot-created articles in Italian, $P(\mathrm{bot}) = 0.06$. 
Moreover, for Bulgarian (bg) Wikipedia (similar trends observed for Malay (ms) and Afrikaans (af)), $P(\mathrm{bot}|o) = 0.51$, which is more than 4 times the overall fraction of bot-created articles in Bulgarian, $P(\mathrm{bot}) = 0.12$. 
This finding could point to undesired artifacts emanating from the use of semi-automatic tools for content creation.

Considering the \textbf{gender} of biography articles, we find that articles about women are over-represented among orphans. 
Overall, we know that between 15-20\% of biographies are about women\footnote{\url{https://humaniki.wmcloud.org/}}. 
However, among orphan articles, we find a much higher percentage of biographies about women. 
For example, in English (en) Wikipedia $P(\mathrm{woman}|o) = 0.29$, which is much larger than the overall fraction of women biographies in English, $P(\mathrm{woman}) = 0.19$. 
An even more extreme example is Catalan (ca) Wikipedia with $P(\mathrm{woman}|o) = 0.42$, while $P(\mathrm{woman}) = 0.20$. 
This shows that biography articles about women are disproportionately more likely to be orphan articles.

Next, \textbf{high-quality} articles are under-represented among orphans, and thus, articles with lower quality are more likely to be orphans across almost all Wikipedia language versions. Moreover, while newer articles (\textbf{age}) are slightly over-represented among orphans, the effect size is relatively small. Finally, all \textbf{topics} are equally represented among orphans; the only exception is `History and Society' which is, on average, substantially underrepresented among orphans.

At this juncture, it is important to note that establishing causality is neither the focus nor the intent of the aforementioned analysis, which solely reveals correlations between different article features and their existence among orphans.

\section{Visibility of Orphan Articles}
\label{sec:visibility-orphans}
In this section, we investigate the visibility of orphan articles to readers navigating Wikipedia. 
By definition, we know that \textit{structurally} they are less visible within Wikipedia because there are no incoming links pointing to orphans from other articles.
Here, we want to assess to which degree this also holds \textit{functionally}, \ie, is this also reflected in orphan articles receiving fewer pageviews?

\subsection{Analyzing Correlations}
As a first step, we compare the pageviews received by orphan articles with that of non-orphan articles, and find that orphans receive substantially fewer pageviews than non-orphans (Fig.~\ref{fig:visibility_orphans_vs_non-orphans}). 
Specifically, due to the long tail in the distribution of pageviews for articles, we compare the mean of the logarithm of the pageviews between the two groups. 
Averaged across all Wikipedia language versions, we find that the mean for non-orphans is twice as high as the mean for orphans. 
This indicates that orphans are, on average, less visible and less visited than non-orphan articles. 
However, this observation is only a correlation, \ie, we cannot conclude that the number of pageviews is lower \textit{because} the articles are orphans.

\begin{figure*}[t]
    \centering
    \includegraphics[width=0.9\linewidth]{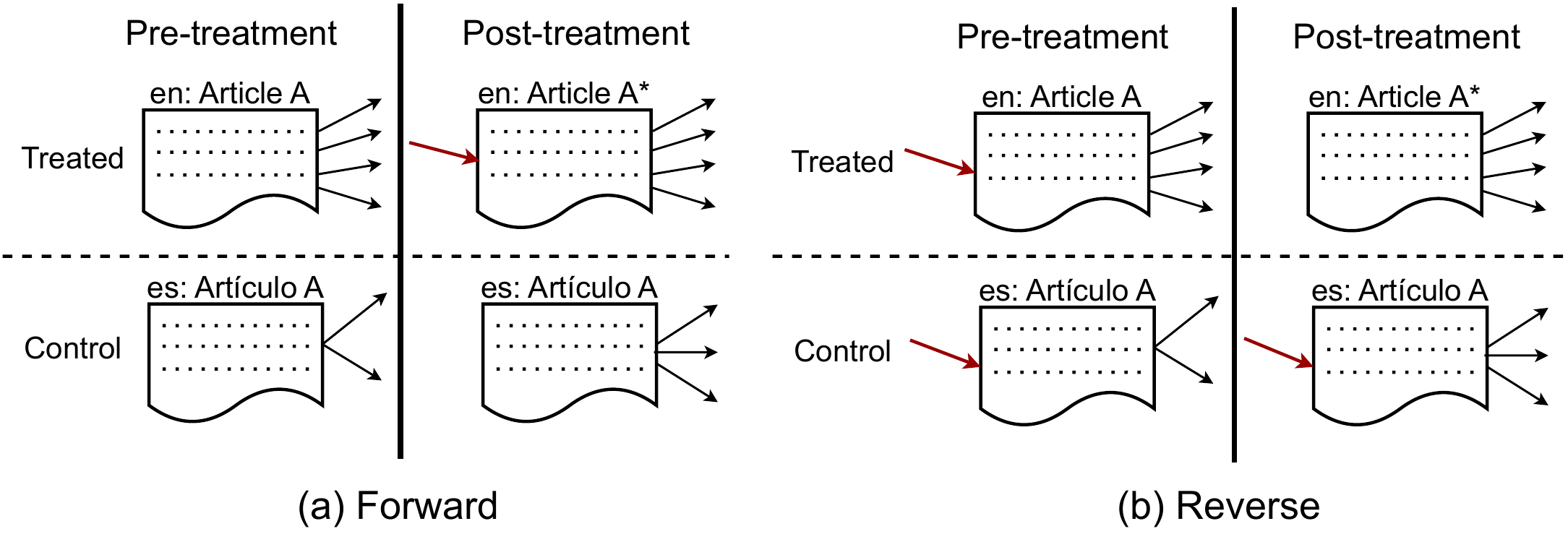}
    \caption{A pictorial representation of the quasi-experiment: (a) Forward: an article that receives a new incoming link (denoted in red font) is considered as treated, whereas the same article in another language that does not receive any new incoming links is considered as control; (b) Reverse: an article that loses an incoming link is considered as treated, whereas the same article in another language that does not lose any incoming links is considered as control.}
    \label{fig:quasi_experiment_setup}
\end{figure*}

\subsection{Establishing Causality}
\label{subsec:causality}
In order to establish a causal link that fewer pageviews are a result of an article being an orphan, we conduct a quasi-experiment~\cite{rosenbaum2017observation} and use difference-in-differences~\cite{angrist_mostly_2008}, a widely used causal inference method, to study the change in the number of pageviews for those orphans that were de-orphanized.

\xhdr{Setting up treatment-control groups}
To setup the quasi-experiment, we follow the process portrayed in Fig.~\ref{fig:quasi_experiment_setup}a. As treatment group, we consider all articles that were orphans in the monthly snapshot of October 2022 but at some point in the following month, \ie, November 2022, received a de-orphanizing incoming link (\cf \textit{de-orphanizing-links data} in Sec.~\ref{sec:data}) so they were not orphans anymore. 
For each de-orphanized (treated) article, we consider the same article, albeit in a different Wikipedia language version in which it remained an orphan, as control. In this way, a given orphan article gets de-orphanized (treated) in one language but remains an orphan (control) in another. 
The motivation to match on the same article is to construct a control group that is as similar to the treatment group as possible, thereby accounting for potential confounding effects due to often fast shifts in attention to specific topics or current events. As a potential limitation, this setup assumes an absence of language-specific shifts in attention, which we discuss in detail in Sec.\ref{subsec:limitations}.

The aforementioned process yields 36,707 treated-control article pairs across 192 language versions (no article was de-orphanized in the remainder 127 versions). 

\xhdr{Difference-in-Differences (DiD)} We use the following DiD model to estimate the effect of de-orphanization on article visibility by comparing the aforementioned treatment-control groups three months before (August--October 2022) and after (December 2022--February 2023) the treatment.
\begin{equation}
\label{eq:vanilla_did}
Y_{it} = \beta_0 + \beta_1\mathrm{deorph}_i + \beta_2\mathrm{after}_t + \beta_3\mathrm{deorph}_i\mathrm{after}_t + \varepsilon_{it}, 
\end{equation}
where $Y_{it}$ is the logarithm of the number of pageviews received by article $i$ in the month $t$, $\mathrm{deorph}_i$ indicates whether article $i$ was de-orphanized or not, $\mathrm{after}_t$ indicates whether the month $t$ is before or after the treatment month, and $\varepsilon_{it}$ is the error term. The coefficient $\beta_3$ denotes the causal effect of article de-orphanization on its visibility measured by the number of pageviews. We extend the aforementioned DiD model to (1) estimate language-specific treatment effect by adding language as a categorical variable into the model, and (2) estimate month-specific treatment effect by transforming $\mathrm{after}_t$ from a binary to a categorical variable.

\xhdr{Results}
The DiD model described in Eq.~\ref{eq:vanilla_did} yields a statistically significant overall increase of $6.5\%$ ($p<10^{-10}$) in the number of pageviews for articles de-orphanized in November 2022. 
Next, we estimate the treatment effect for $120$ language versions in which at least $30$ articles were de-orphanized. While the treatment effect differs across Wikipedia language versions, we find a statistically significant ($p<0.05$) increase for $25$ whereas a decrease for $8$ language versions (the effects were not significant for the remainder $87$ languages), respectively (Fig.~\ref{fig:lang_specific_treatment_effect}). 
The largest increase can be observed for Norwegian (nn) and Haitian Creole (ht), whereas the largest decrease is observed for Cebuano (ceb).

\begin{figure*}[t]
\centering
\includegraphics[width=0.99\linewidth]{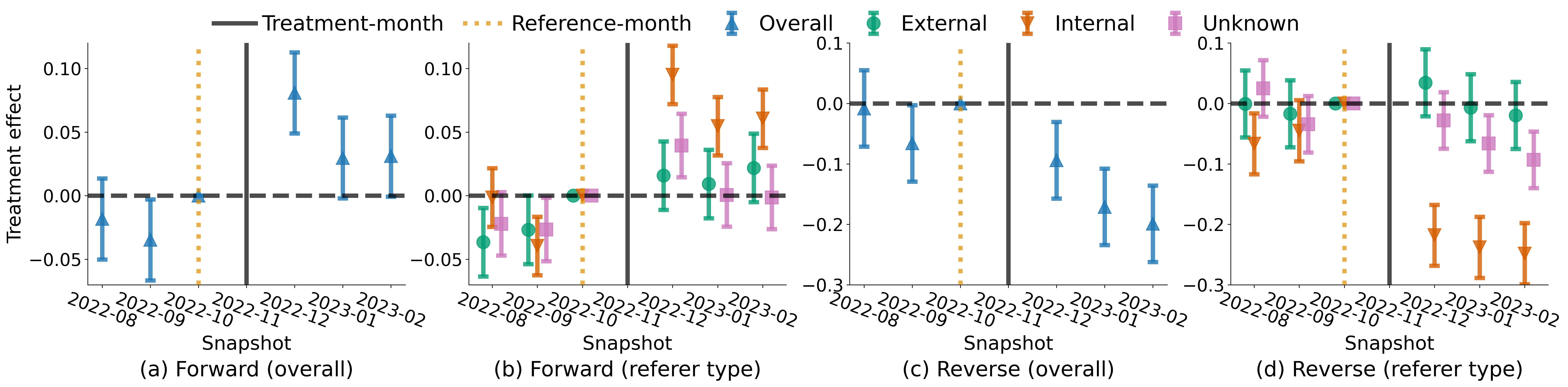}
\caption{Per-month DiD treatment effect with 95\% CIs for the (a)-(b) forward and (c)-(d) reverse setup considering November 2022 as the treatment month.}
\label{fig:treatment_effect_2022-11}
\end{figure*}

Moving ahead, we estimate the month-specific treatment effect (Fig.~\ref{fig:treatment_effect_2022-11}a). It is important to point out the following: (1) we observe a statistically significant ($p<0.001$) positive DiD effect ($7.8\%$) immediately after de-orphanization, (2) the positive effect is persistent for the entire post-treatment duration, and (3) the pre-treatment difference is statistically indistinguishable from $0$, suggesting that our quasi-experimental setup generates treatment and control groups that portray similar behavior prior to de-orphanization, while also providing some evidence in favor of the existence of parallel-trends pre-treatment. Moreover, we obtain qualitatively similar findings if other months are chosen as treatment months (Supplementary Fig.~\ref{fig:treatment_effect_robustness}). 
Overall, the aforementioned points highlight the robustness of our findings.

Finally, we find that the increase in pageviews for the treatment group is, indeed, mostly driven by readers using the newly added incoming links (Fig.~\ref{fig:treatment_effect_2022-11}b). 
For this, we stratify the number of pageviews with respect to their referrer (internal: via a Wikipedia link, external: from an external website or external search engine, and unknown: missing referrer). 
Fitting a separate DiD model for each case yields statistically significant ($p<0.001$) effect sizes only for pageviews with internal referrer.

Note that the vast majority ($80\%$) of de-orphanized articles received exactly one incoming link (Supplementary Fig.~\ref{fig:deorph_indegree_distribution}). Thus, the reported treatment effects in all the aforementioned analyses can be approximately attributed to be emanating from a single link.

\xhdr{Inverting the treatment}
The previous analysis provides causal evidence that adding incoming links to orphans leads to an increase in the number of pageviews. 
An alternative explanation could argue that causality works in the opposite direction, \ie, an increase in the number of pageviews could have lead to the added incoming link because the increase in attention will make it more likely that an editor will encounter and make edits related to the articles.
In order to rule out this alternative explanation, we analyze the inverse process: the treatment group comprises articles that are orphanized whereas the control group comprises articles that remain non-orphans (Fig.~\ref{fig:quasi_experiment_setup}b). 
Considering November 2022 as the treatment month, this process yields 12,560 treated-control article pairs across 121 language versions (no article was orphanized in the remainder 198 versions).

Our DiD model reveals that treated articles experience a statistically significant reduction of $13\%$ ($p<10^{-10}$) in the number of pageviews. 
Further, similar to the forward setup, (1) the pre-treatment difference is statistically indistinguishable from $0$, (2) the reduction is persistent for the entire post-treatment duration (Fig.~\ref{fig:treatment_effect_2022-11}c), and (3) the reduction is prevalent only for pageviews with internal referrer (Fig.~\ref{fig:treatment_effect_2022-11}d). 

It is important to highlight that the causal direction is less contentious in this setup: it is very unlikely that a decrease in pageviews would cause editor activity involving the removal of the corresponding link we considered as treatment. 
Overall, this analysis provides further confidence for establishing the causal direction that adding incoming links to orphan articles leads to an increase in the number of pageviews.

\section{De-orphanization in Practice}
\label{sec:challenges-deorphanization}
In this section, we assess the current state of organic de-orphanization, demonstrate the challenges faced by Wikipedia editors in de-orphanizing articles, and provide potential solutions for developing automated de-orphanizing tools.

\begin{figure*}[t]
    \centering
    \subfloat[Orphans \vs De-orphanization]{
        \includegraphics[width=0.24\linewidth]{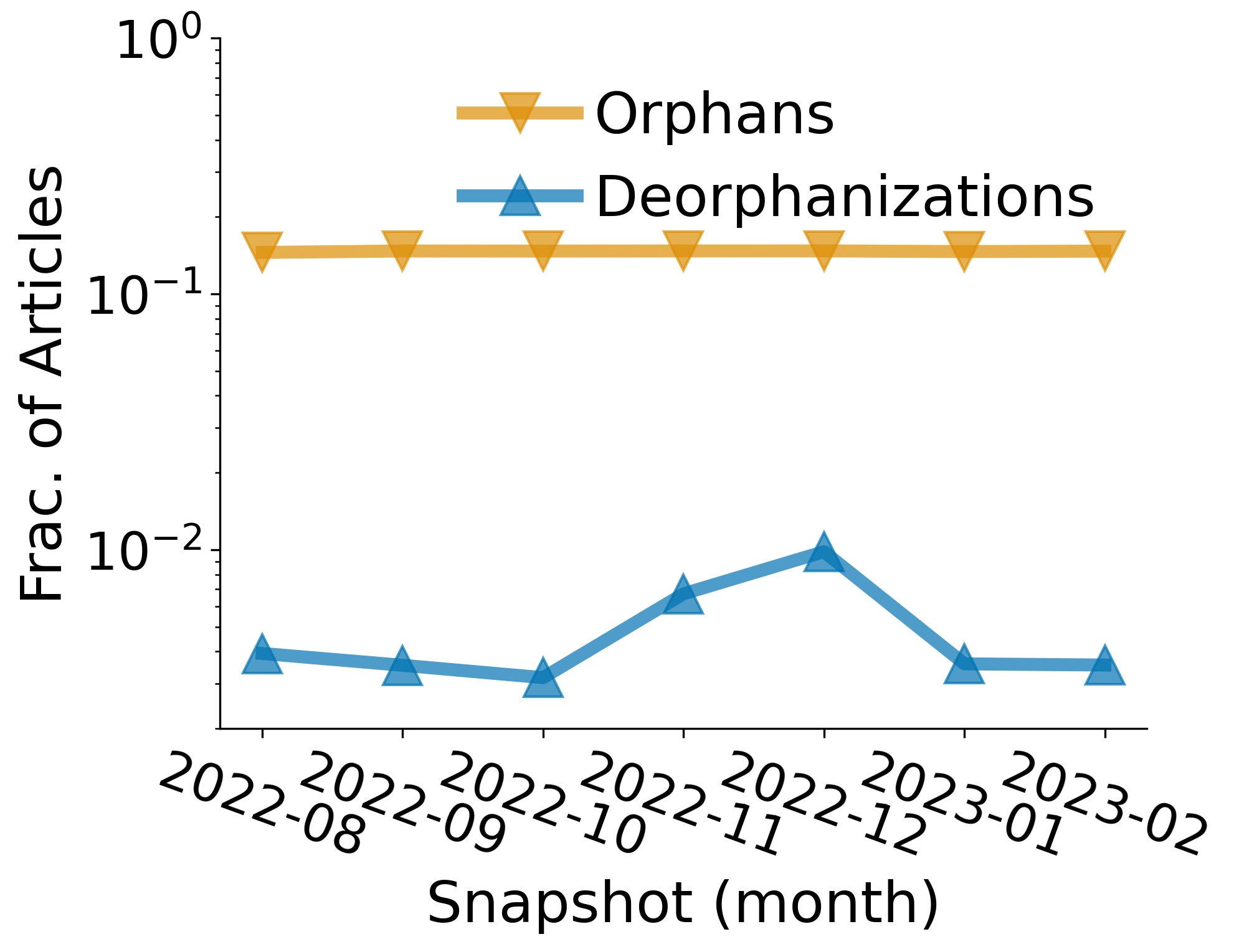}
        \label{fig:rate_of_orphans_and_deorphanization}
    }
    \subfloat[De-orphanizations per Wiki]{
        \includegraphics[width=0.24\linewidth]{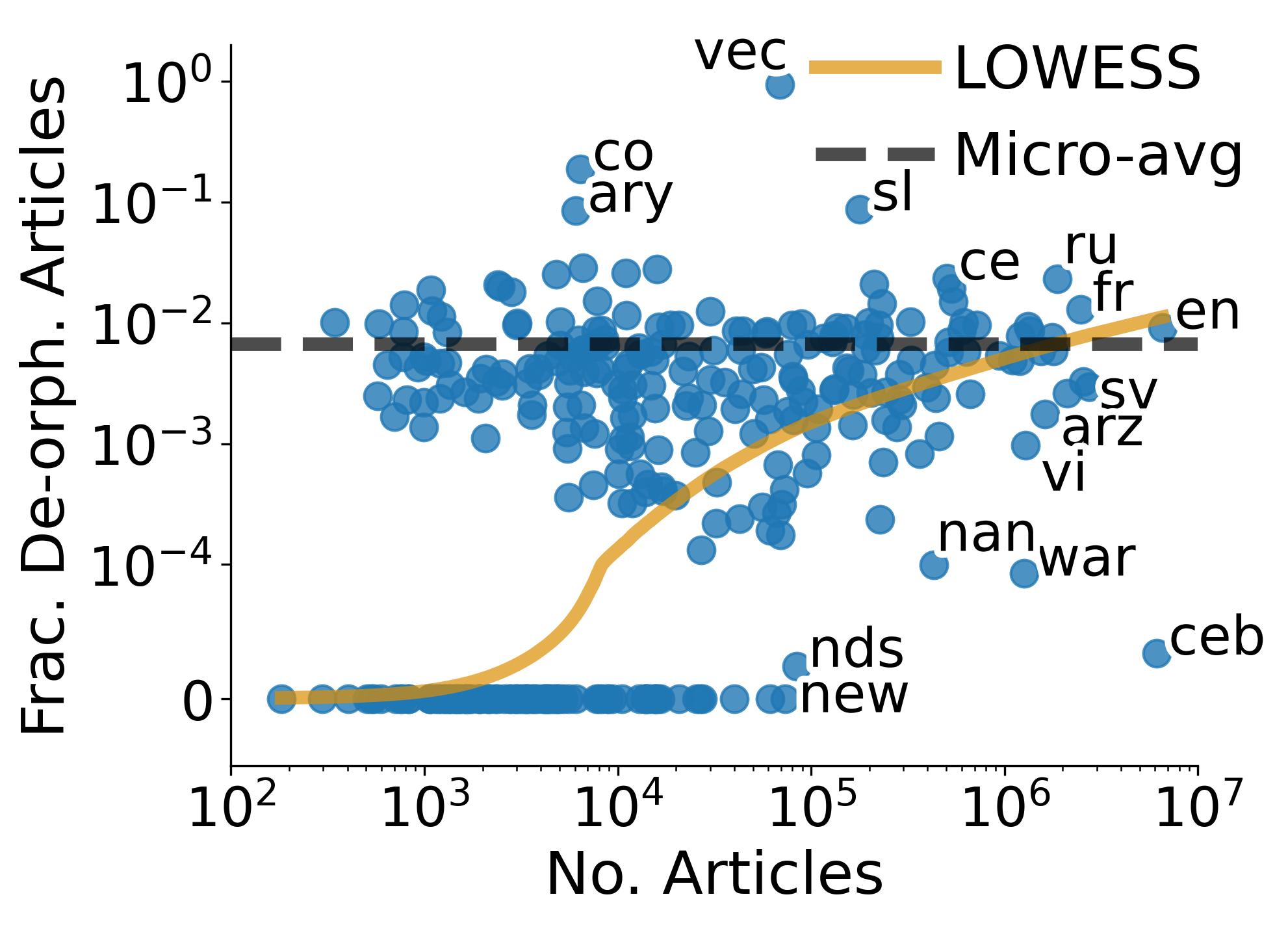}
        \label{fig:deorphanized_per_wiki}
    }
    \subfloat[Findlink]{
        \includegraphics[width=0.24\linewidth]{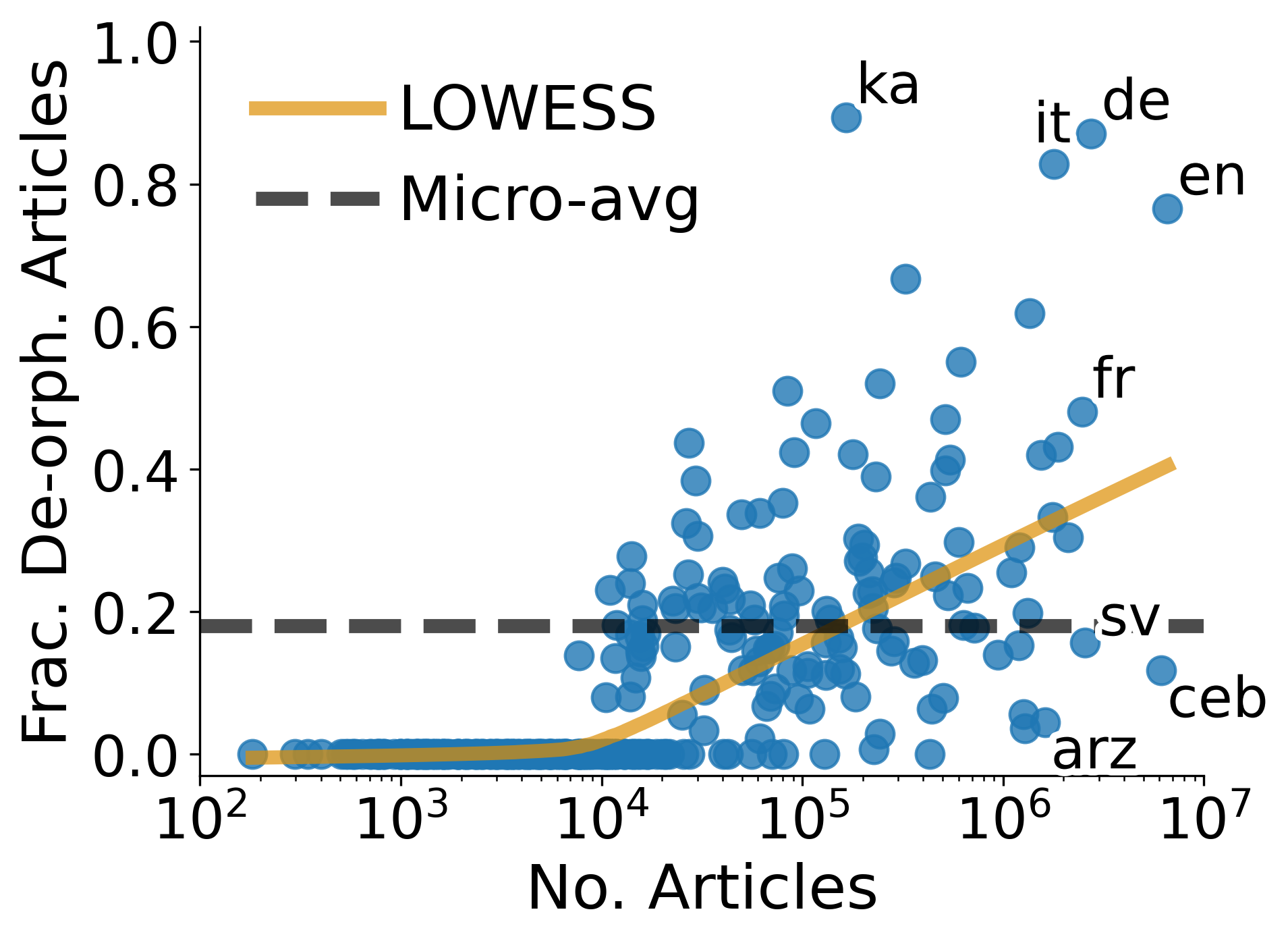}
        \label{fig:findlink}
    }
    \subfloat[Cross-lingual]{
        \includegraphics[width=0.24\linewidth]{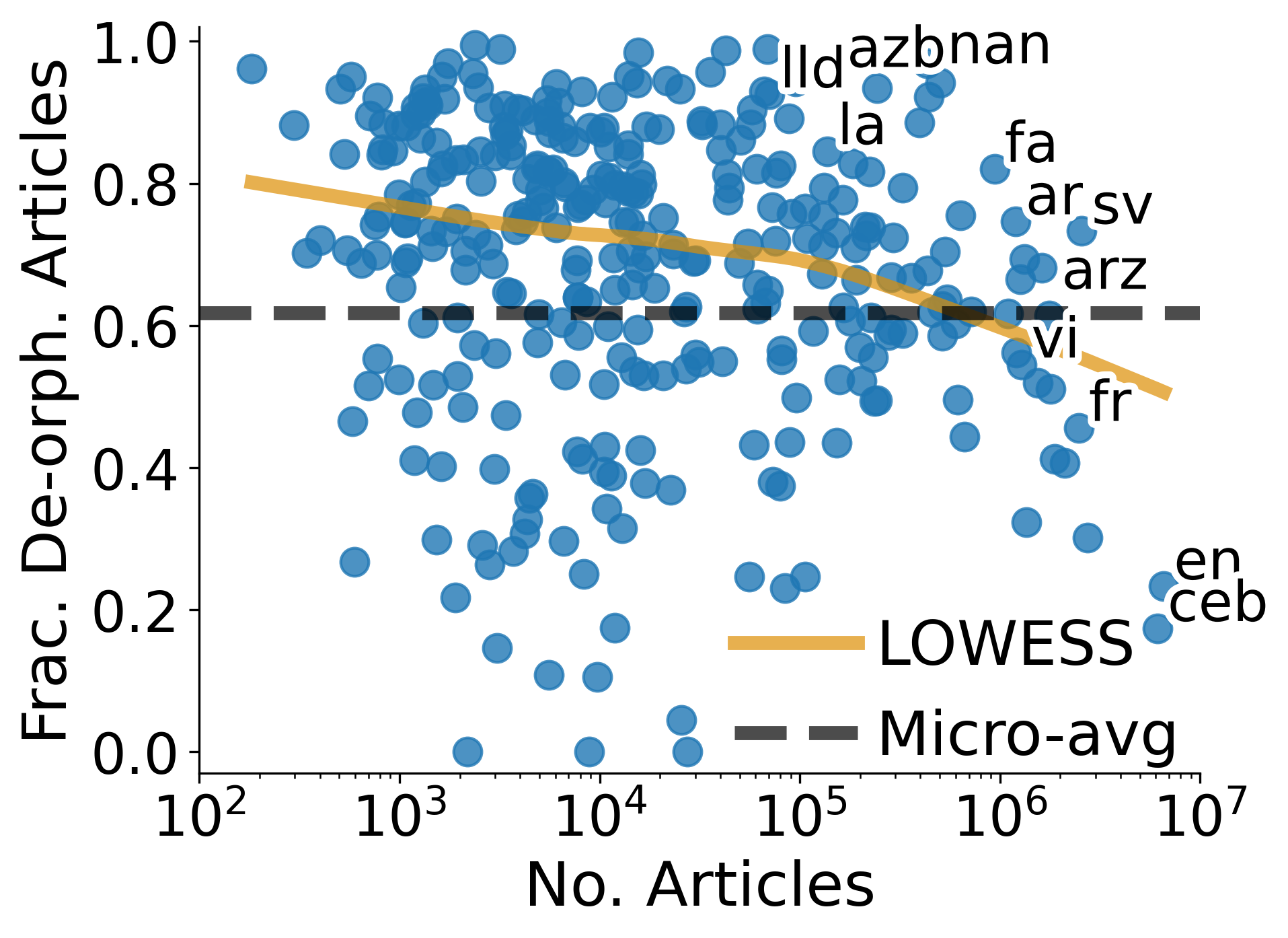}
        \label{fig:link_translation}
    }
    \caption{Analyzing (a)-(b) the current state of de-orphanization, and the fraction of orphans that can be potentially de-orphanized using (c) Findlink, and (d) Cross-lingual approaches across all Wikipedia language versions.}
    \label{fig:challenges}
\end{figure*}

\xhdr{Current state} Wikipedia editors have developed different approaches for de-orphanizing articles such as through marking orphan articles via maintenance templates or coordination via WikiProjects.
While these efforts, on average, facilitate de-orphanization of around 35K articles per month, in comparison to the overall fraction of orphans, the rate of de-orphanization is more than an order of magnitude smaller at around 0.5\% per month (Fig.~\ref{fig:rate_of_orphans_and_deorphanization}). 
With this rate, it would take more than 20 years to work through the backlog of currently existing orphans.
However, with the addition of new content, new orphans are also created such that the overall fraction of orphans remains approximately constant despite the continuous efforts by editors.

We observe some variation in the rate of de-orphanization across Wikipedia language versions (Fig.~\ref{fig:deorphanized_per_wiki}). 
There are only few Wikipedia languages that exceed a rate of 1\% in de-orphanization.
In contrast, there are 94 languages with no de-orphanizations at all.
Taking into account that the latter is more common for smaller language versions, using a LOWESS regression fit we find an overall positive correlation between the size of the language version and the rate of de-orphanization.

\xhdr{Challenges} These observations raise the question about potential reasons for the low rates of de-orphanization.
In \Secref{sec:characterizing-orphans}, we showed that the number of orphan articles is typically much larger than the number of dead-end articles.
This suggests that adding new incoming links is a more difficult task than adding new outgoing links to articles. 
While the latter can be easily added by editing the respective article itself, for adding new incoming links the workflow for editors is more complex because one has to first identify other articles where a link to the orphan articles can be inserted. 

To help editors in this task, the maintenance templates to mark orphan articles suggests the use of the findlink tool~\cite{findlink} to identify suitable candidates. 
Findlink is a community-developed tool that tries to locate unlinked mentions of the specified article title in other articles via a relatively simple text-based search.
However, this approach often yields very few candidates for orphans because the title of the orphan article does not yet appear as a potential mention in the text of any other article. Moreover, findlink works well only for very large language versions, such as English (en), German (de), and Italian (it). 
In fact, the performance is substantially low for smaller language versions (Fig.~\ref{fig:findlink}) with no candidates returned whatsoever for 190 language versions. 
Overall, this approach yields at least one candidate only for 1.6M (18\%) out of 8.8M orphans. 
From this, we conclude that available tools such as findlink are struggling to support editors in identifying candidates for de-orphanization.

\xhdr{Potential solution} Inspired by the success of content translation approaches in Wikipedia~\cite{wulczyn2016growing}, we test whether cross-lingual approaches could be a useful signal to identify candidates for de-orphanization.
The hypothesis is that for an orphan article $a$ in a Wikipedia language $w$, the same article might not be an orphan in another Wikipedia language version $w'\neq w$, thereby possessing an incoming link from article $s$ to $a$ in $w'$. 
If such an article $s$ already exists in the Wikipedia language version $w$, we have identified a natural candidate for a new link from $s$ to $a$ to de-orphanize $a$ in $w$.
These link candidates can generally be considered of high precision because they have already been vetted by one or more communities of editors.
We find that this approach could provide suggestions for a vast majority of orphan articles (Fig.~\ref{fig:link_translation}).
Overall, for 5.5M (62\%) out of 8.8M orphans, this approach yields at least one link candidate for de-orphanization, which already exists in at least one other Wikipedia language version.
In many cases (3.2M), we could actually identify not just one but 10 or more different (incoming) link candidates per orphan article.
While this heuristic is similarly effective across almost all Wikipedia languages, where other languages generally contain link candidates for more than half of the orphan articles, the effectiveness is slightly better for smaller Wikipedia languages. 
In fact, an outlier seems to be English Wikipedia with only 23\% but even this amounts to link candidates for more than 68K orphan articles. We have developed a tool based on this heuristic, which is publicly available at~\url{https://linkrec.toolforge.org/}.

\section{Discussion}
\label{sec:discussion}
\subsection{Summary of Findings}
\xhdr{Many orphans}
The number of orphan articles is surprisingly large: 8.8M ($14.7\%$) out of 60M articles do not have any incoming links. 
This observation is not limited to only a few or small Wikipedia language versions, rather for more than 100 Wikipedia language versions the percentage of orphans is above $30\%$, including Egyptian Arabic ($78\%$) and Vietnamese ($50\%$), which are among the 20 largest Wikipedia language versions. 
In comparison, the number of dead-end articles, \ie, articles without any outgoing links, is very low across all languages (less than $0.5\%$). 
We find that orphan articles are negatively correlated with being: (1) of higher quality and (2) being about the topic of history and society, while possessing a slight positive association with being newer. 
More importantly, we showed that orphan articles encode structural biases: biography articles about women are substantially more common among orphans than expected from their overall frequency.

\xhdr{Lack of visibility}
Orphan articles have, in general, fewer pageviews than non-orphan articles. 
We find causal evidence that orphan articles that were de-orphanized by editors receive a statistically significant increase in the number of pageviews. 
Specifically, we found that this increase is mainly driven by internally-referred pageviews from other Wikipedia articles which contain a link to the de-orphanized article.

\xhdr{Challenges for editors}
The rate of organic de-orphanization is alarmingly low. 
For the snapshots we considered, editors added new incoming links to $\sim$35K orphan articles.
While this constitutes an impressive effort by the community, at that rate it would take approximately 20 years to de-orphanize all orphan articles (assuming no newly created orphan articles).
One hypothesis is that existing tools do not support editors in addressing this issue effectively. 
For example, FindLink (the tool suggested to editors in the orphans maintenance template) generally does not yield many results for orphan articles, especially for smaller languages. 
However, our results show that an orphan article in one language is not always an orphan in other languages. 
This suggests that we can develop an approach for identifying articles from which to link to orphans via link translation. 
Our results shows that this could be effective for 5.5M ($62\%$) orphan articles.

\subsection{Implications and Broader Impact}
\xhdr{Maintenance vs content creation}
While there exists a plethora of efforts to build methods and tools for mitigating content gaps: content translation~\cite{content_translation} to create new content, entity linking~\cite{eigenthemes,gerlach2021multilingual,nelight,lrec_candgen} to ground concepts in knowledge bases, entity alignment~\cite{vldb_ea,Sun2020ABS} to enrich knowledge graphs, \etc, there exists very little support for maintaining the created content.
An important aspect of maintenance work is to integrate new articles into the hyperlink network of Wikipedia. 
While it does not necessarily add new content, it is crucial for the visibility of these articles.
Adding incoming links to articles is also more difficult than adding content to (or creating) the article itself, since it requires editing other articles. 
In fact, it has been shown that community-organized campaigns such as Art{+}Feminism are very successful at improving the content of articles about women; but are less successful at increasing the structural visibility of articles by, \eg, adding new inlinks~\cite{langrock2022gender}. To this end, this work focuses on improving the structural visibility of articles by adding inlinks to orphan articles.

\xhdr{Supporting editors}
Our insights demonstrate the need to support editors to address the issue of orphan articles.
This could be achieved by developing machine-learning models that could generate suggested edits in a machine-in-the-loop approach~\cite{gerlach2021multilingual}.
Such approaches have been shown to be effective for generating outgoing links in the context of structured tasks for newcomer editors~\cite{add_a_link}. 
While the main focus for the latter was the action of the edit itself, extending this framework to adding new inlinks would, therefore, increase the value of the added links.
 
\xhdr{Cross-lingual approaches} 
Our analysis demonstrates the potential of cross-lingual approaches for building well-founded solutions to address the issue of orphan articles.
These cross-lingual approaches not only yield a scalable and robust signal but also have the main advantage that derived models are easily interpretable for editors (\eg editor communities from other Wikipedia languages have already vetted the information).
This is in line with previous work on content translation in Wikipedia.

\xhdr{Knowledge gaps}
Orphan articles provide a more nuanced approach to measuring knowledge gaps in Wikimedia projects~\cite{redi2020taxonomy}.
While the most common approach is to count the number of articles for, \eg, biography articles of different genders, it has been pointed out that this should be complemented by other aspects; most notably the quality and the visibility of articles~\cite{gap_metrics}. 
The current work provides a starting point to systematically operationalize knowledge gaps in terms of their visibility through orphan articles.

\subsection{Limitations and Future work}
\label{subsec:limitations}

While constructing treatment-control groups by matching on the same article (albeit in a different language version) is a powerful way of accounting for most potential confounders, as acknowledged in Sec.~\ref{subsec:causality}, one subtle limitation in this setup is the assumption that different languages portray similar trends in attention shifts for a fixed article. Specifically, one or more language versions may portray an increase in attention for an article (or a topic), which eventually could lead to that article being de-orphanized (treated) and thus, acting as a confounder. Moreover, for a given article (or topic), the expertise of the editor community may also greatly vary across language versions, and thus, an article could be de-orphanized in a given language primarily because of the existence of editor expertise. 
That said, the aforementioned limitations are unlikely to simultaneously impact both the forward (de-orphanization) and reverse (orphanization) setups (Sec.~\ref{subsec:causality}). Overall, considering the two setups in conjunction alleviates most limitations that could impact causal inference.

We showed that pageviews to de-orphanized articles increase significantly. 
One open question is whether these are ``additional'' pageviews or whether this simply corresponds to a shift of pageviews  from other articles (i.e. ``cannibalization''). 
Similarly, does the number of different readers who access these articles also increase?
These questions are difficult to assess not only due to privacy restrictions of the data on readership to Wikipedia articles, but also the fluctuations in overall access volume to Wikipedia in order to disentangle potentially competing articles. Note that we showed an increase in pageviews when considering ``organic'' de-orphanziations performed by editors. It remains an open question whether ``artificial'' de-orphanizations resulting from suggestions to editors via link translation would result in a similar impact, however, it is an interesting question in its own right and constitutes as future work.

While cross-lingual approaches to Wikipedia content contain a rich signal to extend content coverage, there are some caveats.
One challenge is that translations might not meet certain quality standards. 
In an extreme example, it was recently found that nearly half of Scots Wikipedia was created by someone who does not speak Scots~\cite{harrison2020what}.
One additional concern is to avoid ``language imperialism'' and take into account the cultural context~\cite{miquel2018wikipedia}, \ie, different Wikipedia versions cover the same topics differently~\cite{hecht2010tower}.

The preferential attachment model is one of the best-known models to understand observed properties of real-world networks (including Wikipedia~\cite{capocci2006preferential}), \eg, with respect to their degree distributions~\cite{newman2018networks}. 
However, within this framework networks typically yield a large fraction of nodes with in-degree 0 (\ie orphans). 
While this might be natural in many contexts such as the World Wide Web, it is an undesirable property for Wikipedia due to the lack of visibility for much of its existing content. 
This leads to the question of alternative network formation processes that do not lead to large number nodes with in-degree 0; and how to apply this in the context of Wikipedia's communities.

\subsection{Ethical Considerations}
\label{subsec:ethics}
In our opinion, this work has no major ethical considerations. All the datasets and resources used in this work are publicly available and do not contain any private or sensitive information about Wikipedia readers. Moreover, all the findings are based on analyses conducted at an aggregate level, and thus, no individual-level inferences can be drawn. Finally, we took utmost care to distinguish claims establishing causality from those that present non-causal findings, and thus, we do not foresee any negative media impact, especially around misrepresentation of findings, emanating from this research. We confirm that we have read and abide by the AAAI code of conduct.

\section{Conclusion}
\label{sec:conclusion}
Our work constitutes the first characterization of orphan articles as the dark matter of Wikipedia:
a surprisingly large fraction of articles across all 319 language versions of Wikipedia is de-facto invisible to readers of Wikipedia when navigating the hyperlink network. 
Our analysis not only reveals the existence of a causal link between the addition of incoming links to orphans and an increase in their visibility in terms of the number of pageviews, but also demonstrates the need to develop automated tools to support editors in addressing this issue, \eg, via cross-lingual approaches.
The latter would further help address structural biases related to (the lack of) visibility of articles about, \eg, women in the context of the gender gap.
Overall, our results provide a new perspective on the challenges and costs of maintenance associated with the constant creation of new content. 

\setcounter{figure}{0}
\setcounter{table}{0}
\setcounter{section}{0}
\setcounter{subsection}{0}
\makeatletter
\renewcommand{\thefigure}{S\arabic{figure}}
\renewcommand{\thetable}{S\arabic{table}}

\begin{figure*}[!htb]
    \centering
    \includegraphics[width=0.93\linewidth]{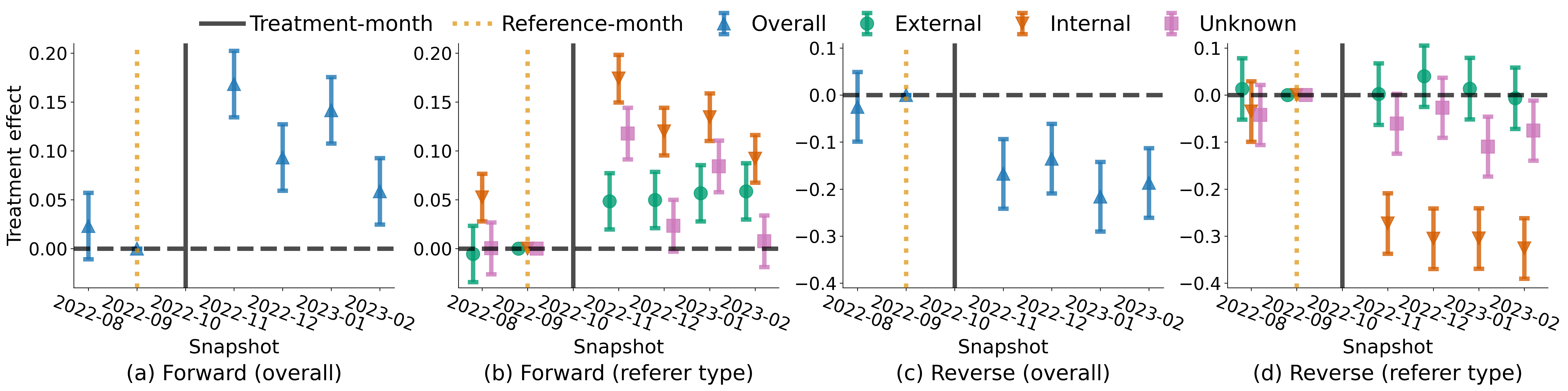}\\
    \includegraphics[width=0.93\linewidth]{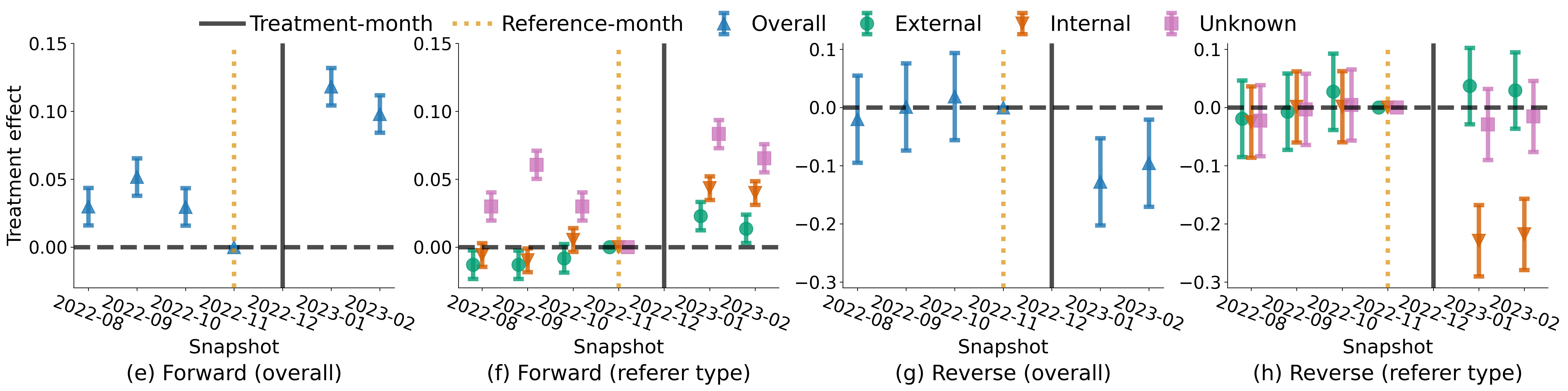}
    \caption{Per-month DiD treatment effect with 95\% CIs for the forward and reverse setup considering October 2022 (top) and December 2022 (bottom) as the treatment month.}
    \label{fig:treatment_effect_robustness}
\end{figure*}

\begin{figure}[!htb]
    \centering
    \includegraphics[width=0.7\linewidth]{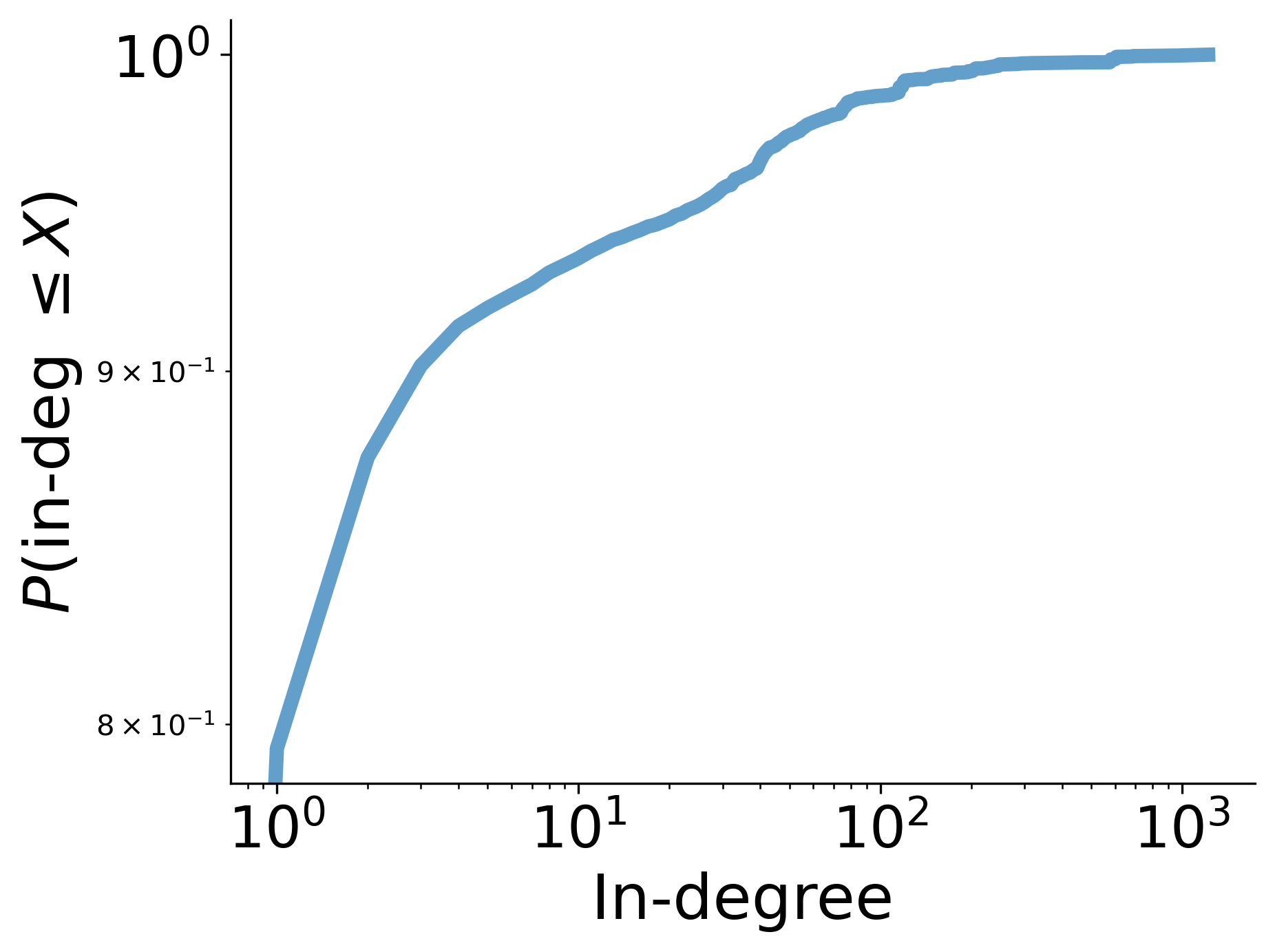}
    \caption{Cumulative distribution function of the number of added incoming links via organic de-orphanization across all Wikipedia language versions in November 2022.}
    \label{fig:deorph_indegree_distribution}
\end{figure}

\section*{Acknowledgements}
We would like to thank Leila Zia, Miriam Redi, Isaac Johnson, Marko \v{C}uljak, Veniamin Veselovsky, Guiseppe Russo, and Manoel Horta Ribeiro for insightful discussions as well reviewing an initial draft of this paper. 
West’s lab is partly supported by grants from Swiss National Science Foundation (200021\_185043), Swiss Data Science Center (P22\_08), H2020 (952215), Microsoft Swiss Joint Research Center, and Google. We also gratefully acknowledge generous gifts from Facebook, Google, and Microsoft.

\appendix

\section{Additional Related Work}
\label{app:rwork}

\xhdr{Article quality models in Wikipedia}
There are several publicly available models that aim to automatically assess the quality of articles in Wikipedia such as ORES~\cite{ores}. 
These models assess the quality based on features extracted from the content available in the specific articles.
Typically they do not consider the number of incoming links~\cite{halfaker2020ores,lewoniewski2019multilingual,warncke2013tell}, though this has been suggested in some works~\cite{dalip2009automatic,anderka2012breakdown,deruvo2015analysing}.

\xhdr{Reader navigation}
Wikipedia’s hyperlinks are crucial for readers’ navigation between articles~\cite{wikinav}. 
While the majority of pageviews originate from an external search engine such as Google (48\%),  38\% of pageviews are referred from other Wikipedia articles (i.e. an internal referrer)~\cite{piccardi2023large}.
When considering reading sessions (i.e. combining sequentially visited articles by the same reader), most readers visit only a single article (68–72\%) and thus do not use hyperlinks. 
However, the distribution of the number of visited articles shows a long tail with tens of millions of reading sessions consisting of 10 or more pageviews~\cite{piccardi2022going}. 
Interestingly, it was found that a substantial fraction of readers use an external search engine to navigate between articles despite the availability of a corresponding hyperlink in Wikipedia. 
Such phenomena have been the subject of different efforts to sketch the interdependence between Wikipedia and search engines~\cite{mcmahon2017substantial,vincent2019measuring} as well as other online platforms more generally~\cite{vincent2018examining}.

\bibliography{orphans}

\end{document}